# PHYSICS-INFORMED MACHINE LEARNING FOR SEISMIC RESPONSE PREDICTION OF NONLINEAR STEEL MOMENT RESISTING FRAME STRUCTURES

*R. Bailey Bond[a], Pu Ren[b], Jerome F. Hajjar[a,1], and Hao Sun[c,1]*

[a] Department of Civil and Environmental Engineering, Northeastern University, Boston, MA 02115, USA
[b] Machine Learning and Analytics Group, Lawrence Berkeley National Lab, Berkeley, CA 94720, USA
[c] Gaoling School of Artificial Intelligence, Renmin University of China, Beijing, 100872, China

**Abstract:** There is growing interest in using machine learning (ML) methods for structural metamodeling due to the substantial computational cost of traditional simulations. Purely data-driven strategies often face limitations in model robustness, interpretability, and dependency on extensive data. To address these challenges, this paper introduces a novel physics-informed machine learning (PiML) method that integrates scientific principles and physical laws into deep neural networks to model seismic responses of nonlinear structures. The approach constrains the ML model's solution space within known physical bounds through three main features: dimensionality reduction via combined model order reduction and wavelet analysis, long short-term memory (LSTM) networks, and Newton's second law. Dimensionality reduction addresses structural systems' redundancy and boosts efficiency while extracting essential features through wavelet analysis. LSTM networks capture temporal dependencies for accurate time-series predictions. Manipulating the equation of motion helps learn system nonlinearities and confines solutions within physically interpretable results. These attributes allow for model training with sparse data, enhancing accuracy, interpretability, and robustness. Furthermore, a dataset of archetype steel moment resistant frames under seismic loading, available in the DesignSafe-CI Database [1], is considered for evaluation. The resulting metamodel handles complex data better than existing physics-guided LSTM models and outperforms other non-physics data-driven networks.

Keywords: physics-informed machine learning, seismic response prediction, nonlinear dynamics, fragility analysis, steel moment resisting frame

## 1. INTRODUCTION

Modern structural analysis models for complex and nonlinear engineering systems heavily rely on computational methods to solve numerical simulations. The finite element method (FEM) [2], with applications in fluid flow, heat transfer, electromagnetic potential, structural analysis, and many other domains, is one of the most common simulation-based methods used in dynamic model creation and nonlinear response history analysis (NRHA). In the last decade, computational power has grown synchronously with complex engineering problems with intricate geometries, nonlinear hysteretic material behavior, nonstationary response variables, and under stochastic spatiotemporal loading [3]. Despite the growing supply of high-performance/cloud computing clusters and facilities, demands of growing FEM complexity for NRHA are still prohibited by computational power and run time. Furthermore, optimization techniques (e.g. topology optimization [4], particle swarm optimization [5,6] etc.), and epistemic uncertainties of external loading require numerous simulations (e.g., Monte Carlo simulations [7,8] and incremental dynamic analysis (IDA) [9,10] of nonlinear structural systems for fragility and reliability analysis) which can exponentially increase the computational demand.

[1]Corresponding Authors

Structural metamodeling can supplement more expensive computer analyses, facilitate multi-objective optimization and design exploration, and possess faster run times compared to traditional structural modeling for repetitive loading [11–13]. Metamodeling has taken many forms, including statistical analysis [14–19], response surface methodologies (RSM) [20–22], and kriging [23–26]. The model output associated with these metamodeling techniques offers varying degrees of response detail. Some techniques produce course variables (e.g., static engineering design parameters (EDPs), like peak displacements, drifts, etc.). Other methods keep detailed time-varying information for more in-depth analysis, capturing frequency, duration, and energy content of the time series response. Metamodels only capturing EDPs often cannot quantify cumulative damage metrics of structural components and sometimes rely on simplified assumptions (e.g., the generalized linear regression models assumes a normal distribution [14,15]).

Recently, neural networks (NN) and machine learning (ML) have been shown to supplement existing metamodeling approaches or used on their own, creating a new class of structural metamodeling all together and can be classified as data-driven approaches [13,27–31]. Particularly, pioneering studies have applied convolutional neural networks (CNNs), recurrent neural networks (RNNs), and long short-term memory (LSTM) networks to project the time history seismic responses of civil structures [32–34]. Oh et al. [35] utilized a CNN architecture to predict building displacement responses from ground motion acceleration measurements. Perez-Ramirez et al. [36] harnessed a nonlinear autoregressive exogenous model to train an RNN, predicting response time histories for a 1:20 scaled, 38-story high-rise building structure exposed to seismic excitations and ambient vibrations. Similarly, an RNN-based methodology, utilizing gated recurrent units in tandem with ensemble learning, was proposed by Zhang et al. [37] to scrutinize the time-variant uncertainty in structural response. Ahmed et al. [38] employed a stacked LSTM model to evaluate the seismic damage states of frame buildings and non-ductile bridges, demonstrating its superior accuracy and efficiency in contrast to conventional LSTM models. Moreover, Zhang et al. [39] harnessed stacked GM sequences as inputs, constructing an LSTM model to predict displacement response histories of nonlinear steel structures under seismic loads. Soleimani-Babakamali et al. [40] devised an encoder-decoder architecture with LSTM and CNN layers for probabilistic seismic demand analysis for a building inventory, propagating structural uncertainties into seismic demand models. Kundu et al. [41] introduced an LSTM-based algorithm to quantify seismic response uncertainties by addressing the stochastic nature of dynamic loads and uncertainties in structural system parameters. Li et al. [42] introduced a model order reduction technique with subsampling and wavelet transformations with a standard LSTM network for seismic response prediction with synthetic ground motion data and matching structural response of high-fidelity structures. Torky and Ohno [43] introduced a fusion ConvLSTM-LSTM model with data order reduction techniques on various data representations for real-time system identification. Ning et al. [30] compare three ML models (e.g., LSTM, WaveNet, and CNN) to model the structural response of three structures. While these models have demonstrated sufficient predictions of structural behavior on their respective datasets, due to the nature of *black box* learning, these data-driven approaches require large synthetic or augmented datasets to reach desired levels of accuracy. Additionally, in many cases, the interpretability of a *black box* model is questionable and has been found to produce responses that are outside of physical possibilities [44].

Purely mechanics driven models (e.g., FEM and NRHA) aim to explain natural phenomena through mathematics utilizing Newtonian physics to resolve high fidelity structural equilibrium and compatibility which has been refined for centuries and validated with real world data. Pure data-driven techniques, relatively new in development, have been shown to capture highly nonlinear and complex behavior, poses fast inferencing capabilities, and are computationally efficient after training, but require large datasets with unique examples and generally perform poorly for cases outside of the training dataset [30,42]. Hybrid modeling aims to combine the benefits of physic-based and data-driven approach [45,46]. Hybrid modeling is particularly effective when data is limited and generally is lower fidelity compared to purely data-driven modeling approaches. However, it still maintains statistical significance and relies less on human engineered features compared to models based entirely on physics. For example, constraints on the CNN and RNN (e.g., LSTM) models can inform physical insights into the training process to guide models to physically relevant solutions and deal with small datasets common in structural earthquake engineering [32]. Zhang et al. [45] embedded physics into a one-dimensional CNN model to predict seismic responses of three datasets, including sensor measurements from a real structure, for model training and validation. Zhang et al. [46] expanded on this work by incorporating multiple LSTM networks to model hysteretic and state variable response separately while utilizing the equation of motion in model development. Wang and Wu [47] utilize physics in a knowledge-enhanced deep learning algorithm to simulate the wind-induced linear/nonlinear structural dynamic response. The synthesis of the temporality of RNNs was employed by Sadeghi Eshkevari et al. [48] in a physics-based RNN model to estimate the dynamic response of systems subjected to seismic ground motions. Although these methods produce detailed reconstructions of the time history response of their training data, the structures being modeled are relatively simple (e.g., single-degree of freedom structures, or shear-type multi-degree of freedom structures with numerical based hysteresis, or structures that are less than 3 stories) and might not be capable of expressing complex behavior of modeled material and geometric nonlinearities for a range of building heights, including tall buildings.

To this end, a physics-informed machine learning (PiML) metamodel is presented in this work to balance nonlinear relationships of NN parameters and Newtonian physical models. Model order reduction techniques are taken from the data-driven literature to enhance the training of hybrid model for time series prediction of special steel moment resisting frame (SMRF) structures under dynamic seismic loading. LSTM networks are chosen for their superior performance in sequence-to-sequence modeling and established performance for metamodeling full sequence time series. A physics-informed network was chosen to constrain the solution space to physical bounds and allow the model to be trained on less data with a real measured ground motion dataset. The presented PiML metamodel is shown to predict structural response of complex archetype building frames, previously unaccounted for in the full time series response prediction of PiML metamodel literature. Additionally, the presented model is shown to be trained and tested with a dataset of force-response pairs from the commonly used

Miranda ground motion set [49] opposed to synthetic ground motions common in many of the above studies.

## 2. METHEDOLOGY

*2.1 Problem Definition*

A structural metamodel is used to capture underlying nonlinear behaviors of archetypal buildings from input-output (e.g., force-response) data. These models are trained with high-fidelity NRHA simulations, like FEM, or measured sensing data of a structure, to reproduce the structural behavior (e.g., time history response, peak EDP, cumulative damage states, etc.). For illustrative purposes, consider the response of a nonlinear multi-degree-of-freedom (MDOF) structure under dynamic loading governed by the equation of motion (EOM):

$$\mathbf{M}\mathbf{u}_{tt}(t) + \mathbf{C}\mathbf{u}_t(t) + \mathbf{f}_s(t) = -\mathbf{M}\boldsymbol{\Gamma}\mathbf{a}_g(t), \quad t \in \tau \qquad (1)$$

$$\mathbf{f}_s(t) = \lambda \mathbf{K}\mathbf{u}(t) + (1-\lambda)\mathbf{K}\mathbf{r}(t) \qquad (2)$$

where $\mathbf{M}, \mathbf{C}, \mathbf{K}$ are the mass, damping, and stiffness matrices; $\mathbf{u}_{tt}, \mathbf{u}_t, \mathbf{u} \in \mathbb{R}^n \times \tau$ are the acceleration, velocity, and displacement vectors of dimensionality $n$, with $\tau$ time-steps, relative to the ground; $\boldsymbol{\Gamma}$ is the force distribution vector over the system; $\mathbf{a}_g$ is the time series earthquake ground acceleration vector; $\mathbf{f}_s(t)$ is the restoring force function; $\lambda$ is the ratio of post-yield to pre-yield stiffness; and $\mathbf{r}(t)$ is the non-observable hysteretic parameter. The restoring force function is complex and of high order and can be thought of more generally as $\mathbf{f}_s(t) = G(\mathbf{u}(t), \mathbf{r}(t))$, where $G$ is an unknown latent function. The metamodel aims at finding a parsimonious solution of $G$, mapping the mass normalized seismic input, $\boldsymbol{\Gamma}\mathbf{a}_g(t)$, to the nonlinear structural response, $\mathbf{u}(t)$ and its derivatives. Eqn. (3) is formulated by rearranging and mass normalizing Eqn. (1), where $\boldsymbol{\xi}_d$ is a mass normalized damping matrix.

$$\mathbf{g}(t) = \mathbf{M}^{-1}\mathbf{f}_s(t) \approx -\boldsymbol{\Gamma}\mathbf{a}_g(t) - \mathbf{u}_{tt}(t) - \boldsymbol{\xi}_d \mathbf{u}_t(t) \qquad (3)$$

In NRHA of structures with large nonlinearities influenced by complex dynamic loading, fast estimation of $\mathbf{g}(t)$ is of significant interest.

*2.2 Model Order Reduction*

For many reasonable archetypal structures, the dimensionality of the state variable response (e.g., $n$) often exceeds hundreds or more. Such high dimensionality poses challenges in terms of training stability, computational demands, and often results in an unnecessary level of precision for metamodeling applications. To surmount these challenges, model order reduction techniques have been developed to identify distinguishing characteristics of the input order dimension, thereby reducing the size of the dataspace, and substantially mitigating the computational time associated with ML model training.

One classical model order reduction method is to introduce a reduced basis of generalized coordinates using static condensation of the initial FEM. This approach is grounded in the recognition that DOFs with mass dominate the dynamic behavior of structures, and thus, by selectively isolating these DOFs, the resulting reduced-order model can capture the fundamental response characteristics. This strategy leads to enhanced computational efficiency, capturing the primary motion of the structure, and simplifying the model creation and validation with physical intuition. However, it is important to acknowledge that by exclusively considering DOFs with mass, certain complex behaviors arising from higher-order dynamics or localized effects (associated with non-mass DOFs) may be overlooked. Therefore, the applicability of this approach should be evaluated based on specific requirements, desired accuracy levels, and available computational resources. The aim of PiML metamodeling in this approach is not to capture smaller-scale details of the structure, but to understand the behavior on a more global scale. Thus, in the ensuing examples, the focus will be on selectively isolating and learning the responses of DOFs with mass through the PiML learning process.

*2.3 Wavelet Analysis*

In dynamic structural analysis problems, force-response pairs generally consist of large amounts of time steps. Employing ML modeling on datasets with large sequence lengths, such as ground motion time series, will increase computational costs and could cause memory problems. Therefore, it can be beneficial to reduce the temporal dimension.

Wavelet approximations offer many advantages for this task [50]. Wavelets are mathematical functions that divide data into different frequency components, allowing each component to be analyzed with a resolution matched to its scale [50]. They are particularly useful for analyzing signals that have non-stationary or variable characteristics over time [50]. Firstly, by decomposing the signal into different scales and frequencies, wavelets can capture localized features and generalized variations within the dataset [50]. This capability makes wavelet approximations particularly well-suited for capturing the salient dynamics of complex structural models. Secondly, wavelets provide a flexible framework for time-frequency analysis, enabling the representation of both low-frequency trends and high-frequency oscillations, separating true data-structures from noise [50]. This versatility allows for a more comprehensive exploration of the underlying dynamics, facilitating a more accurate and nuanced PiML metamodel. Moreover, wavelet approximations can effectively handle non-stationary and irregularly sampled time series, which are commonly encountered in real-world applications.

The wavelet function, $\psi(t)$ is characterized by its scaling function, $\phi(t)$, effectively replicating a band-pass filter where each scaling level halves the bandwidth. Following the data preprocess methodologies outlined in [42], the dyadic wavelets are adopted here:

$$\psi_s(\tau; t) = 2^{\frac{s}{2}} \psi(2^s t - \tau) \qquad (4)$$

where $s \in \mathbb{Z}$ is the wavelet scaling parameter and $\tau \in \mathbb{Z}$ is the time shift parameter identifying translation of the wavelet. In this context, $\tau$, has constant meaning to Eqn. (1). Among the family of dyadic wavelets, the Daubechies wavelet of order 6 is chosen, similar to [42] and suitable for engineering applications for its compact form and orthogonality. The scaling function is thus defined as:

$$\phi(t) = \sum_\tau c(\tau) \phi(2t - \tau) \qquad (5)$$

where $c(\tau)$ is the scaling coefficient. The wavelet function can then be obtained for all time steps using Eqn. (6) and Eqn. (4):

$$\psi(t) = \sum_\tau (-1)^\tau c(\tau + 1) \phi(2t + \tau) \qquad (6)$$

For a particular time history, $\mathcal{X}(t)$, the wavelet transformation is accomplished by:

$$\mathcal{X}(t) = \sum_\tau W_{s_a,\mathcal{X}}(\tau) \phi_s(\tau; t) + \sum_{s_d=0}^{S} \sum_\tau W_{s_d,\mathcal{X}}(\tau) \psi_{s_d}(\tau; t) \qquad (7)$$

where $W_{s_a,\mathcal{X}}(\tau)$ and $W_{s_d,\mathcal{X}}(\tau) \in \mathbb{R} \times \mathbb{Z}$ are the approximated and detailed coefficients (e.g., from the high-pass and low-pass filters), respectively. By ignoring the detailed component of Eqn. (7) (e.g., the second term in Eqn. (7)) a wavelet approximation of the original signal can be made:

$$\mathcal{X}(t) \approx \sum_\tau W_{s_a,\mathcal{X}}(\tau) \phi_s(\tau; t) \qquad (8)$$

From the approximated coefficients and use of the scaling function, a reconstruction of the original signal can be produced. In the metamodeling context, a wavelet approximation can be applied to the reduced input (ground motion) and reduced output (structural response variables) from the model order reduction parameters of the structure:

$$q_j(t) \approx \sum_\tau W_{s_a,q_j}(\tau) \phi_s(\tau; t) \qquad (9)$$

$$p_j(t) \approx \sum_\tau W_{s_a,p_j}(\tau)\phi_s(\tau;t) \tag{10}$$

where $q_j(t)$ and $p_j(t) \in \mathbb{R} \times \tau$ are the $j$th reduced input and output. Similar to [42], the approximated wavelet coefficients (e.g., $W_{s_a,q_j}(\tau)$ and $W_{s_a,p_j}(\tau)$) are thus directly applied to the PiML metamodel for training and testing.

*2.4 LSTM Network*

In recent years, NNs continue to proliferate in scientific computing [51]. Weighted connections and shifting bias parameters allow various architectures of connected artificial neurons to learn patterns to approximate a generalized form of input data. Additionally, different connections have been developed for different types of data (e.g., numerical, categorical, text, images, and time series) and to solve different tasks (e.g., pattern recognition, natural language processing, temporal forecasting, etc.). A training stage, utilizing training data shifts model parameters until model performance is deemed satisfactory followed by a testing stage, utilizing data previously unseen by the model, with fixed model parameters to validate the model.

A RNN is a class of networks with one set of shareable network parameters; the connections between nodes form a cycle or loop [32]. RNNs are commonly used on time series data for detection, generation, synthesis, and translation tasks. Figure 1(a) shows an RNN loop with block "**L**" where the input $X_\tau$ and output $Y_\tau$ occur at every time step and another connection is made as an output of **L** which is then fed back into the network [52]. This loop is known as a recurrent connection, allowing the network to learn temporal behavior. The network can be unrolled with time (e.g., Figure 1(b) [52]). While training an RNN, the network parameters associated with **L** will update with different training examples. During testing, however, the network parameters in **L** are consistent and do not change with time.

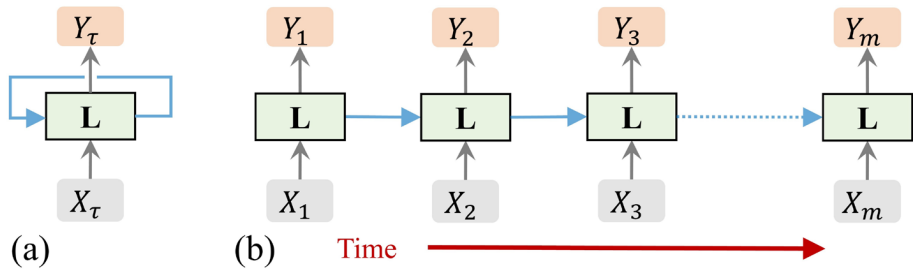

Figure 1: (a) RNN loop; (b) An "unrolled" RNN [52]

Long Short-Term Memory (LSTM) Networks were introduced by Hochreiter and Schmidhuber [34] as an extension to RNNs. The recurrent connection of an LSTM network links the output of a network with itself to give the learning process persistence, utilizing information learned earlier in the sequence data in the prediction of the next time [33,52]. A typical LSTM network consists of an input layer (e.g., time series sequence), hidden layers (e.g., multiple LSTM and fully connected neural network (FCNN) layers), and an output layer (e.g., time series sequence). The term "LSTM" can be applied to the network, the layers in the network, or the specific units that make up the LSTM layer. An input element at time $\tau$ is mapped through the LSTM network to an output element at time $\tau$ for $\tau = [1, m]$ making this network adept to sequence-to-sequence modeling. The input and output sequences must be formatted as three-dimensional tensors; the first dimension is the sequence length and changes with time, the second dimension is the batch size or data object number, and the third dimension consists of the input or output features (a.k.a. the channel or model order).

Figure 2 shows an abbreviated network architecture and individual LSTM unit. The LSTM unit shares weights and biases across the entire temporal space within the layer. Figure 2 (a) shows an "unrolled" LSTM network like Figure 1(b). The main highway of the LSTM unit is termed the cell state, which passes information from previous time steps to future time steps, with minor linear interactions [52]. Four feeder gates interact with the cell state to accomplish the recurrent nature of the LSTM cell and learn complex time-dependent data structures: a forget gate, an input gate, a hyperbolic tangent layer, and an output gate. The input gate and hyperbolic tangent layers filter input activation into the internal cell state, the output gate regulates output activation into the LSTM unit output, and the forget gate effectively throws away unneeded information from the cell state.

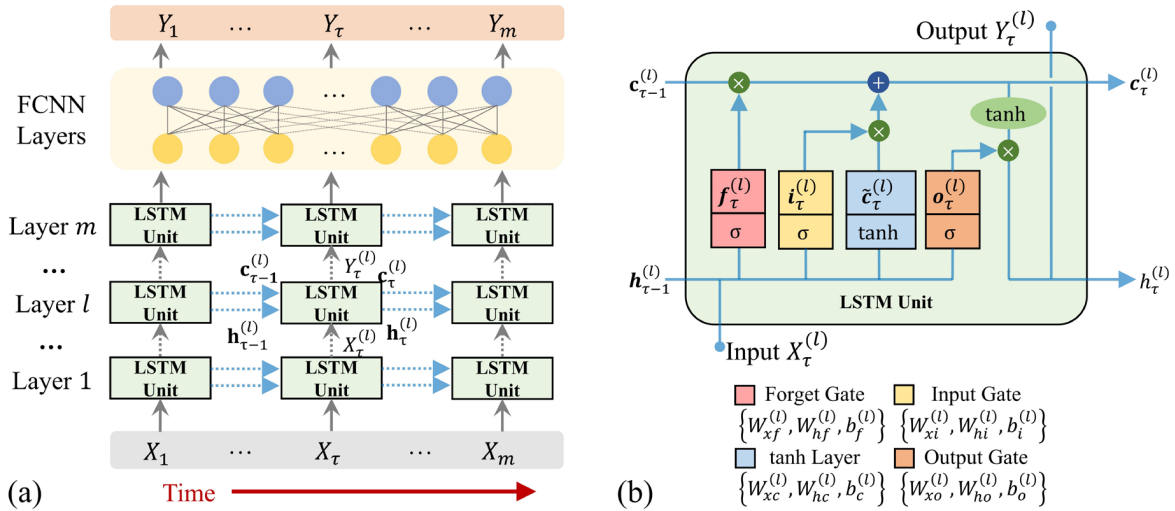

*Figure 2: (a) Architecture of a deep LSTM Network with $m$ LSTM layers and FCNN layers from input sequence $X_n$ to output sequence $Y_n$; (b) single LSTM unit structure of the lth LSTM layer at time $\tau$,*

*including the unit's input $X_\tau^{(l)}$, the unit's output $Y_\tau^{(l)}$, the cell state input and output $\{c_{\tau-1}^{(l)}, c_\tau^{(l)}\}$, the hidden state input and output $\{h_{\tau-1}^{(l)}, h_\tau^{(l)}\}$, and the memory gates variables $\{f_\tau^{(l)}, i_\tau^{(l)}, \tilde{c}_\tau^{(l)}, o_\tau^{(l)}\}$.*

The following equations demonstrate the operations of the feeder gates and their corresponding weights, $W_\alpha^{(l)}$, and biases, $b_\alpha^{(l)}$, (where $\alpha = \{f, i, c, o\}$ corresponding to the forget gate, input gate, tanh layer or output gate, respectively) within the $l^{th}$ LSTM layer. Reference Figure 2 and Equations (11)-(16) for weight and bias interactions from the input as $X_t$, forget gate as $f_t^{(l)}$, input gate as $i_t^{(l)}$, hyperbolic tangent gate as $\tilde{c}_t^{(l)}$, output gate as $o_t^{(l)}$, cell state memory as $c_t^{(l)}$, and the hidden state output as $h_t^{(l)}$ which corresponds to the output as $Y_t$ [46].

$$f_t^{(l)} = \sigma\left(W_f^{(l)} \cdot [h_{t-1}, X_t]^{(l)} + b_f^{(l)}\right) \tag{11}$$

$$i_t^{(l)} = \sigma\left(W_i^{(l)} \cdot [h_{t-1}, X_t]^{(l)} + b_i^{(l)}\right) \tag{12}$$

$$\tilde{c}_t^{(l)} = \tanh\left(W_c^{(l)} \cdot [h_{t-1}, X_t]^{(l)} + b_c^{(l)}\right) \tag{13}$$

$$o_t^{(l)} = \sigma\left(W_o^{(l)} \cdot [h_{t-1}, X_t]^{(l)} + b_o^{(l)}\right) \tag{14}$$

$$c_t^{(l)} = f_t^{(l)} * c_{t-1}^{(l)} + i_t^{(l)} * \tilde{c}_t^{(l)} \tag{15}$$

$$h_t^{(l)} = o_t^{(l)} * \tanh\left(c_t^{(l)}\right) \tag{16}$$

where $\sigma$ is the logistic sigmoid function; tanh is the hyperbolic tangent function; $*$ represents the Hadamard product (element-wise product). A pass through a LSTM layer encodes long-term dependence, allowing important information learned from the beginning of the time series to pervade through to the end and unimportant information to be forgotten. The FCNN layers then decode the last LSTM layer output into the necessary data structure of the desired output space.

A data-driven approach, utilizing a simple LSTM network, is created to compare the performance enhancements of the PiML metamodel. The LSTM network outputs the state variables response of the structure directly; this output is compared to the ground-truth state variables to form the data loss as seen in Figure 3. In this data-driven approach, the loss function only consists of the data loss.

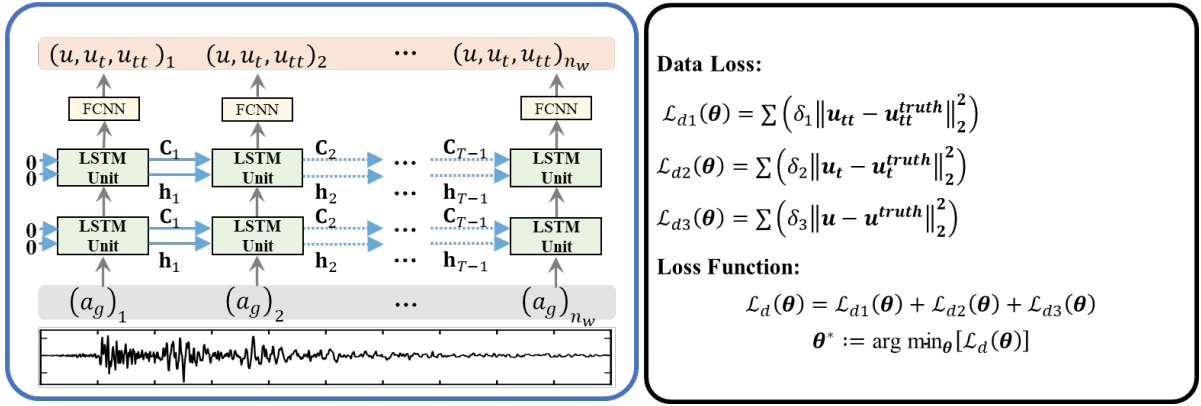

*Figure 3: Simple data-driven LSTM metamodel architecture*

*2.5 Physics-Informed Machine Learning Metamodel Architecture*

The proposed PiML architecture extends the recent utilization of multiple Deep LSTM networks for physics-*informed* metamodeling techniques from [46] and is capable of handling larger nonlinearities and more complex datasets. PiML in the proposed network utilizes four main features: Step 1 - the LSTM network, Step 2 – a wavelet reconstruction of the full time-series, Step 3 – a finite difference filter, and Step 4 – application of Newton's second law (e.g., the EOM) through the nonlinear restoring force variable, $g$. Figure 4 shows the general stages of the PiML metamodeling framework. The unique aspect of this network relies on a physical constraint to the objective function (e.g., loss function) during training of the LSTM network's weights and biases. The LSTM network itself is unchanged with regards to the network architecture used in traditional ML.

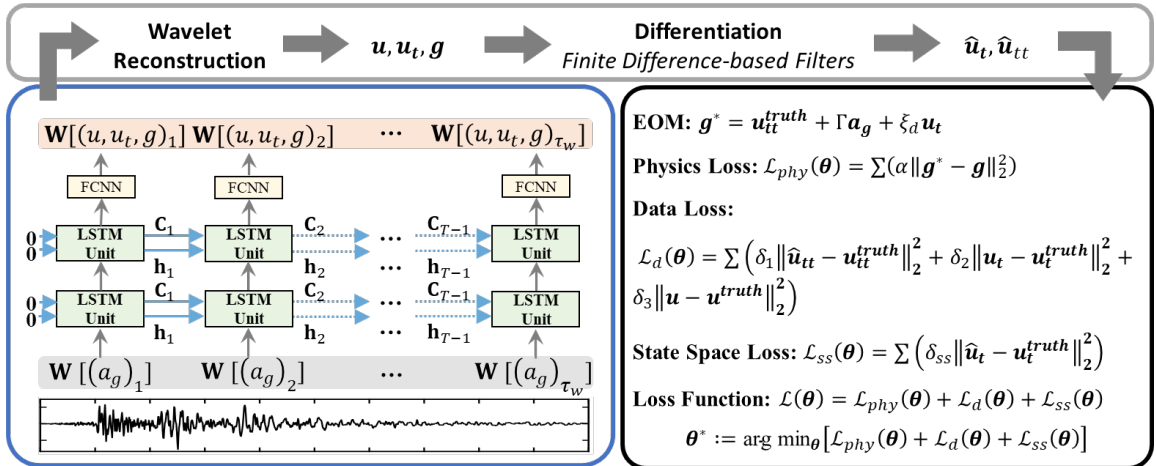

*Figure 4: Schematic architecture of the PiML metamodel*

The first LSTM layer inputs the wavelet coefficients of the control force, and outputs a hidden state, cell state, and output to the corresponding networks blocks. A total of two LSTM layers are used followed by an FCNN consisting of three hidden layers with the hyperbolic

Bond - 11

tangent activation function. The output of the FCNN (e.g., the network output, $\mathbf{W}[\mathbf{u}, \mathbf{u}_t, \mathbf{g}]$) consists of three nodes, corresponding to the wavelet coefficients of the displacement, velocity, and restoring force. The network output is then passed to wavelet reconstruction to produce the full time series of the response variables (e.g., $\mathbf{u}, \mathbf{u}_t, \mathbf{g}$) and then to a finite difference-based filter.

The finite difference-based filter (e.g., a gradient-free convolutional filter) is utilized to represent discrete numerical differentiation [53,54] and calculate a derivative term for both the physics loss and state-space loss. Specifically, the first-order central difference method is considered to approximate the time derivatives based on the learned variables, given by:

$$K_t = \frac{1}{2\delta t}[-1,0,1]. \tag{17}$$

where $\delta t$ is the time interval. It is worth noting that the computation of derivatives for the first and last steps is not directly feasible due to the inherent nature of the central finite difference scheme. Thus, forward, central, and backward differences are combined to get a derivative variable length equal to the input variable length. The displacement, $\mathbf{u}$, passes the filter and outputs the velocity vector, $\hat{\mathbf{u}}_t$. The velocity, $\mathbf{u}_t$, passes the filter and results in the acceleration vector, $\hat{\mathbf{u}}_{tt}$. The differentiation step enforces the network outputs to obey differentiation rules among the state variables, confining the solution space to physical bounds.

A new variable, $\mathbf{g}^*$, is constructed using the training data and network output as described in Equation (18). $g^*$ represents the *pseudo ground truth* restoring force of the structure and will be compared with the $\mathbf{g}$ of the wavelet reconstructed network output. The known structure of the EOM is formulated on the right-hand side of Equation (18), while the unknown and unobservable nonlinear portion is isolated on the left-hand side. Thus, $\mathbf{g}^*$ will iterate to a more optimal representation of the data while constraining the learning of the state variable response.

$$\mathbf{g}^* = \mathbf{u}_{tt}^{truth} + \mathbf{\Gamma} a_g + \xi_d \mathbf{u}_t \tag{18}$$

The loss function (e.g., model error function) is constructed by combining three functions: the residual loss (physics loss), data loss, and state space loss. The residual loss, $\mathcal{L}_r(\theta)$, minimizes L2 norm error between $\mathbf{g}^*$ and $\mathbf{g}$. The data loss, $\mathcal{L}_d(\theta)$, minimizes the L2 norm error between the ground truth of the training state variables, $\mathbf{u}_{tt}^{truth}, \mathbf{u}_t^{truth}, \mathbf{u}^{truth}$ with the differentiation output $\hat{\mathbf{u}}_{tt}$ and the network output $\mathbf{u}_t$ and $\mathbf{u}$, respectively. Finally, the state space loss, $\mathcal{L}_{ss}(\theta)$, minimizes the L2 norm of the network output $\mathbf{u}_t$, and the differentiation output $\hat{\mathbf{u}}_t$. Each of these terms will be weighted to balance their effect on the overall optimization function, $\mathcal{L}(\theta)$. These weights are known as hyperparameters that can change the overall performance of loss minimization. The data loss weights ($\delta_1, \delta_2, \delta_3$) are found according to Equations to (19)-(21). The state space loss weight, $\delta_{ss}$, is set to equal $\delta_2$ because both loss components hold similar magnitudes. The residual loss weight, $\alpha$, is found through a parametric study, but generally to make the loss components contribute to the global loss equally.

$$\delta_1 = 1.0 \tag{19}$$

$$\delta_2 \approx \frac{mean\left(std(u_{tt}^{truth})\right)}{mean\left(std(u_t^{trtuh})\right)} \tag{20}$$

$$\delta_3 \approx \frac{mean\left(std(u_{tt}^{truth})\right)}{mean(std(u^{truth}))} \tag{21}$$

For each training iteration (e.g., each epoch), the loss function, $\mathcal{L}(\theta)$ from Figure 2, will produce an error or gradient value which will be backpropagated through the layers of the network, attributed in some portion to the model parameters, $\theta$ (e.g., LSTM and FCNN weights and biases). These parameters will be updated via stochastic gradient descent and the Adam optimizer [55].

In this application, the LSTM network is thought to be capable of learning order dependence in the response data while the finite difference-based filter and the EOM confine the solution space to realistic results. These features reduce the need for large training sets, relieve overfitting issues, and increase the robustness of the trained model for more reliable prediction. The model order reduction of the training and testing data further speed up the training and reduce redundancy in the response data. Validation and performance examples of the network architecture are shown in the following section.

3. EXAMPLES

In the following examples, nonlinear time-history responses of structures are generated by numerical simulations to compose the *ground-truth* data. This ground-truth data is compared to the structural response generated with the PiML metamodel as well as a response generated with a data-driven LSTM network. The case studies consist of nonlinear special steel moment resisting frames (SMRFs), sourced from the database created by Guan et al. (2020) [56] and downloaded from the DesignSafe-CI Data Depot [1,57]. The total dataset comprised 81 one-story, 149 five-story, 122 nine-story, 78 fourteen-story, and 38 nineteen-story steel moment frames. The parameters considered in the dataset are shown in Table 1.

*Table 1: SMRF Database Design Specifications* [57]

| Category | Parameters | Values considered in archetype design space |
|---|---|---|
| Geometric Configuration | Number of stories | 1, 5, 9, 14, and 19 |
| | Number of bays | 1, 3, and 5 |
| | First story/typical story height | 1.0, 1.5, and 2.0 |
| | Bay width | 20 ft (6.10 m), 30 ft (9.14 m), and 40 ft (12.19 m) |
| | Number of LFRSs | Two in principal direction |
| | Typical story height | 13 ft (3.96 m) |
| Load Information | Floor dead load | 50 psf (2.39 kN/m$^2$), 80 psf (3.83 kN/m$^2$), & 110 psf (5.27 kN/m$^2$) |
| | Roof dead load | 20 psf (0.96 kN/m$^2$), 67.5 psf (3.23 kN/m$^2$), & 115 psf (5.51 kN/m$^2$) |
| | Floor live load | 50 psf (2.39 kN/m$^2$) |
| | Roof live load | 20 psf (0.96 kN/m$^2$) |
| Design conservatism | Allowable drift limit | 2% |
| Steel strength | Yield stress | 50 ksi (345 MPa) |

*3.1 Structural Finite Element Model and Seismic Excitation*

The SMRF design methodology was derived and validated from experimental data from the Applied Technology Council (ATC)-123 project [56–59]. Frames were designed using the equivalent lateral force (ELF) analysis procedure. Seismic story forces were approximated from the fundamental period and calculated with ASCE 7-16. Initially, seismic forces are estimated using the building's fundamental period, in accordance with ASCE 7-16 [60]. Preliminary beam and column sizes are decided with engineering judgment and basic rules of thumb [56], considering accidental torsion. An elastic model is analyzed for gravity and lateral seismic forces, and structural deformations and member forces are obtained. Story drifts are then adjusted and compared against limits (e.g., 2% allowable drift); if they exceed, member sizes are increased and the process is repeated, including stability checks. Further, the structural members are checked for ductility, strength, and stability according to specific standards [60–63]. The commonly used reduce beam section (RBS) connection was chosen, ensuring they meet strength and design criteria. The site conditions matched that in the ATC-123 project [58], specifically site class D and spectral accelerations of = 2.25g and = 0.6g. Finally, the design is refined for practical construction considerations, like using consistent member sizes across floors and accounting for deeper columns in lower stories for splices. Further details on the steel frame archetypes can be found in Guan et al. (2020) [56].

To analyze the steel frames, numerical models were created and assessed using the OpenSees software [64]. The beams and columns were represented using concentrated plastic hinge beam-column elements to simulate the nonlinear material behavior of steel beam-column components. These elements comprised two nonlinear hinges at both ends and a linear elastic beam-column element in the middle. A zero-length rotational spring featuring a modified Ibarra-Medina-Krawinkler (IMK) material [65,66] was employed for the modeling of nonlinear hinges. The possible shear yielding at the SMRF panel zone is considered through panel zone modeling using a combination of elastic elements and zero length rotational springs [56]. Additionally, a leaning column, connected to the frame with truss elements, is used to account for P-Δ effects. The leaning column's hinge is simulated using a zero-length rotational spring that possesses minimal rotational stiffness. For simplicity, the training and testing results of one SMRF model per story level in the dataset (namely 1, 5, 9, 14, and 19 stories, corresponding to building IDs 0, 81, 567, 1053, and 1539 [57]) were chosen to demonstrate the PiML metamodel's performance in the following sections. Figure 5 depicts the structural layouts of example SMRF structure and a detail of the panel zone.

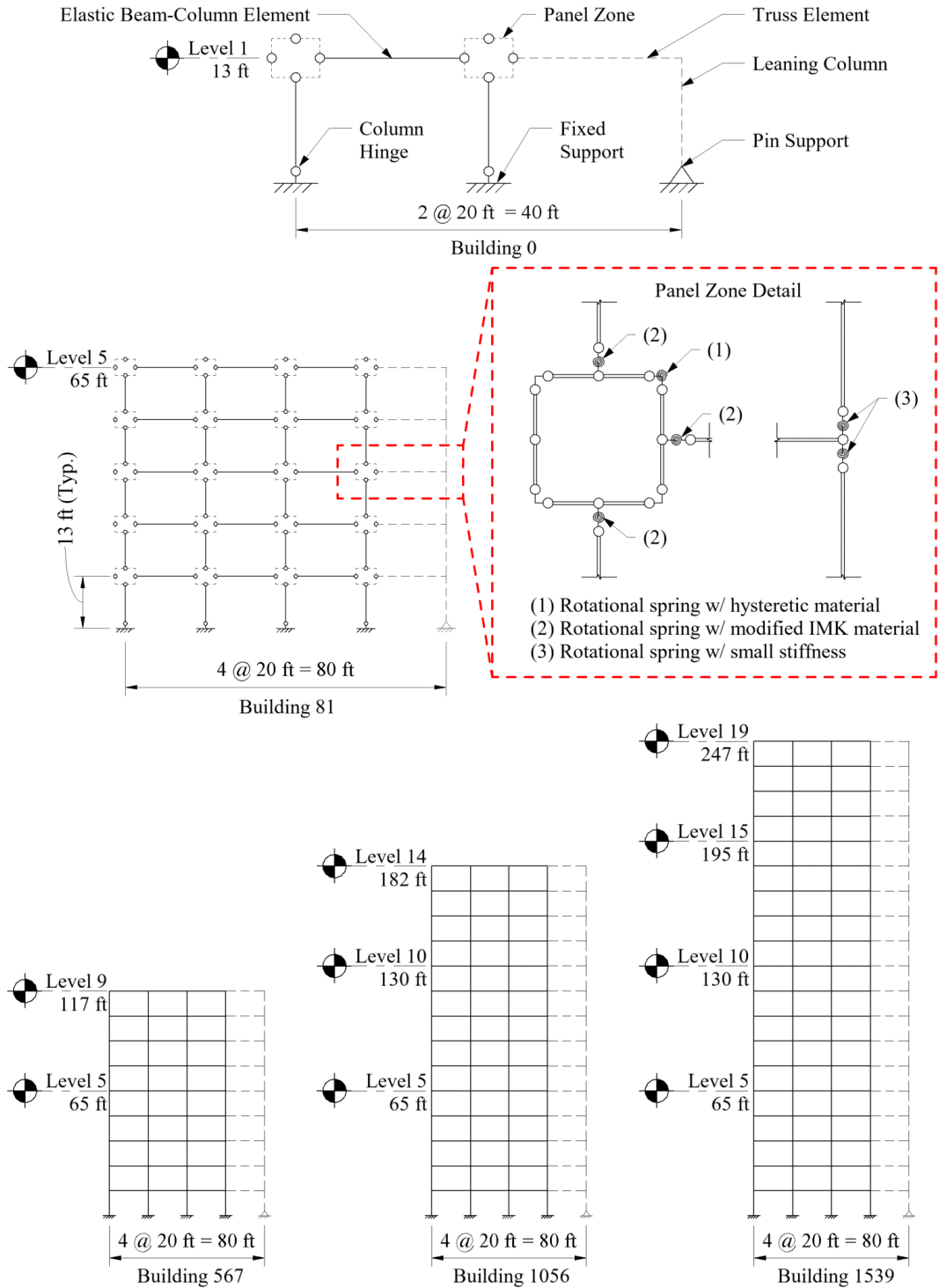

*Figure 5: Five 2D SMRF structural layouts* [56]

A synthetic dataset of nonlinear time-history responses of the structure is generated under the excitation of 240 ground motions obtained from a study by Miranda (1999) [49] to align with the Miranda suite of the SMRF structural models [56,57].

The ground motions were scaled utilizing the FEMA P695 [67] methodology to amplify the targeted training and testing data in which the residual drift is significant and the relative nonlinear hysteretic behavior is large. Record scaling is a two-step process [67]. Step 1: Normalization of records with normalization factor ($NF$): this factor normalizes the ground acceleration time series by their respective peak ground velocities to remove unwarranted variability between records due to inherent differences in event magnitude, distance, source type, and site conditions without eliminating record-to-record variability [67]. Step 2: Anchoring of records with scale factor 1 ($SF_1$): this factor is used to "anchor" the ground acceleration time series by a specific ground motion intensity such that median spectral acceleration of the record set ($S_M$) matches the design level spectral acceleration at the fundamental period of structure ($\hat{S}_S$) [67]. The two layers of scaling are described in Equations (22)-(24).

$$a_g^{scaled}(t) = NF * SF_1 * a_g^{recorded}(t) \tag{22}$$

$$NF = \frac{\text{median}(PGV)}{PGV} \tag{23}$$

$$SF_1 = \frac{S_M}{\hat{S}_S} \tag{24}$$

Due to the anchoring step, each ground motion suite is scaled slightly differently based on structural building properties (e.g., first mode fundamental period). The 5% critically damped response spectra of the unscaled and scaled ground motions are shown in Figure 5 for the 1 story SMRF. For the unscaled ground motions, the peak ground acceleration (PGA) of the records vary from 0.03g to 0.77g with a mean of 0.17g, while the scaled ground motions for the record sets are shown in Table 2. The records were recorded at free fields or the first floor of a low-rise structure at firm sites with shear wave velocities greater than 590 ft/s (180 m/s) from the upper 100 ft (30 m) layer. More details of the ground motion suite can be found in Miranda (1999) [49].

*Table 2: SMRF ground motion suite*

| Building Information | | | Ground Motion Suite Information | | | |
|---|---|---|---|---|---|---|
| Building ID [57] | Number of Stories | First Mode Period [sec] | Average PGA [g] | Min PGA [g] | Max PGA [g] | Std. PGA [g] |
| 0 | 1 | 0.511 | 0.98 | 0.19 | 3.41 | 0.62 |
| 81 | 5 | 1.107 | 0.89 | 0.17 | 3.10 | 0.56 |
| 567 | 9 | 1.547 | 0.69 | 0.14 | 2.40 | 0.44 |
| 1053 | 14 | 1.818 | 0.61 | 0.12 | 2.14 | 0.39 |
| 1539 | 19 | 2.311 | 0.50 | 0.10 | 1.73 | 0.31 |

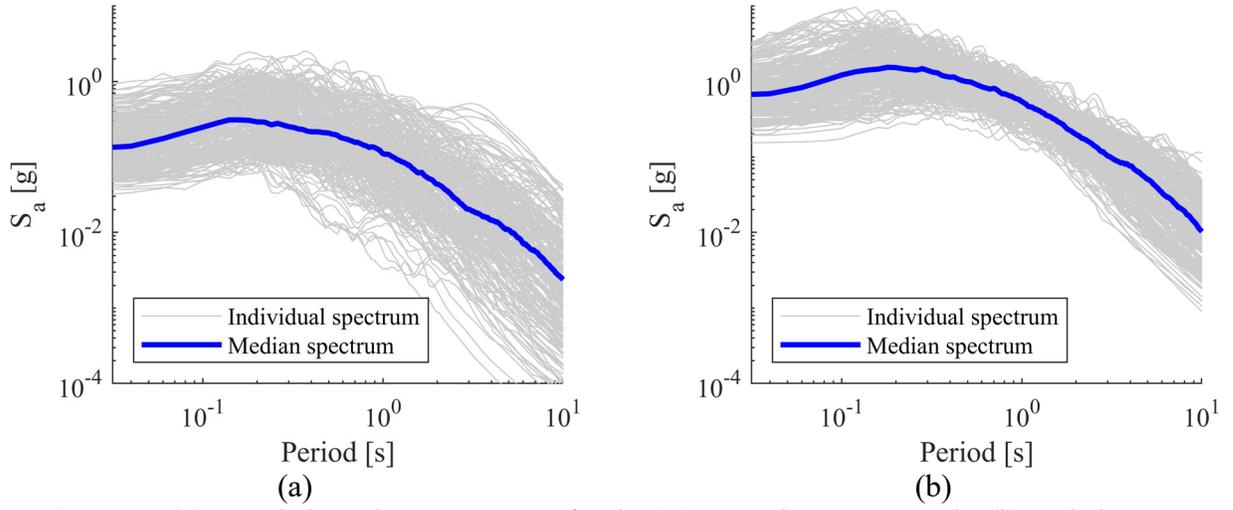
*Figure 6: (a) unscaled acceleration spectra for the 240 ground motion records; (b) scaled acceleration spectra for Building 0.*

## 3.2 PiML Metamodel Training and Testing

The input scaled ground accelerations were combined with the response of the high-fidelity finite element SMRF model to create input/output data pairs. To establish the size for the PiML metamodels' training set, the number of data pairs was initially set at 20 and then progressively increased in increments of 20, until it reached a maximum of 200 pairs. As expected, the model performance enhanced with larger training sets, but in general improvements plateaued at 100 pairs for the PiML metamodels. Therefore, for training, 100 pairs were randomly selected from a total of 240 ground motion and corresponding state variable responses. The remaining 140 pairs were allocated for validation and testing.

So that each channel, or DOF of the structure, contributes equally to the training process, an absolute scaling method is administered such that each value in a channel is normalized between -1 and 1. The scaling is done independently for each channel, based on the maximum absolute value in that channel and the result maintains the sign of the original data. After normalization, a wavelet transformation was applied as described in Section 2.4. Specifically, this transformation used a sixth-order Daubechies wavelet with a scale parameter of s = 2, generating wavelet coefficients used for training and testing.

During the training phase, the loss value, derived in Figure 4, attributes gradients to the network parameters, $\theta$, and are backpropagated through the layers of the network. A parametric analysis was performed to tune the network hyperparameters of the PiML metamodel.

Specifically, a minibatch size of 20, weight decay parameter of 1e-6, and initial learning rate of 2e-2 were chosen, alongside a learning rate decay period of 300 epochs and a multiplicative factor of 0.95. These hyperparameters were uniformly applied across all datasets and proved effective in minimizing the loss function across various scenarios. While data-specific tuning of these hyperparameters may enhance performance for each individual dataset, pursuing data specific tuning is unfeasible within real-world implementations. A hyperbolic tangent activation function was chosen for the FCNN as it outperformed other common activation functions during testing. Additionally, the incorporation of Batch Normalization (Batch Norm.) layers between hidden layers of the FCNN facilitated the attainment of a consistent distribution of activation values throughout the training process. The training was conducted using a computer system equipped with 28 Intel Core i9-10900KF CPUs and NVIDIA GeForce RTX 3090 GPU card, which featured CUDA compatibility, facilitating accelerated deep learning operations. The computer code was implemented in the PyTorch framework [68].

The calibrated (e.g., trained) PiML metamodel, which utilized the network parameters $\theta$ (e.g., weights and biases or $W_\alpha^{(l)}$ and $b_\alpha^{(l)}$) from the minimum validation loss epoch, was used to predict structural state variable responses with unknown and unseen ground motions as input (e.g., the testing set ground motions). The testing data also includes the reference (e.g., ground truth) output state variable response, which can then be compared to the metamodel output to display the performance of the PiML metamodel. Figure 7 shows a typical result (e.g., response data for the high-fidelity model, the PiML metamodel, and the data-driven LSTM network) for each structure tested under one ground motion input. Successful validation implies that the model can adequately predict structural responses under different conditions.

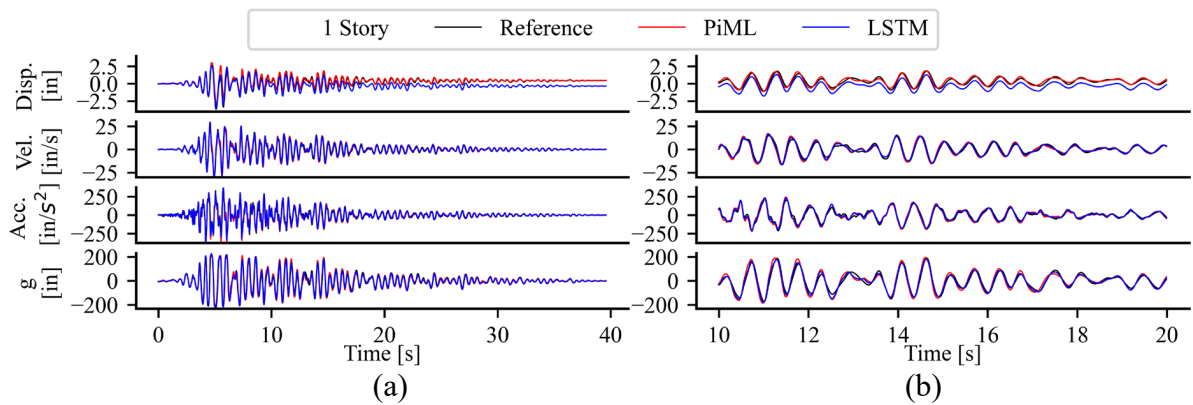

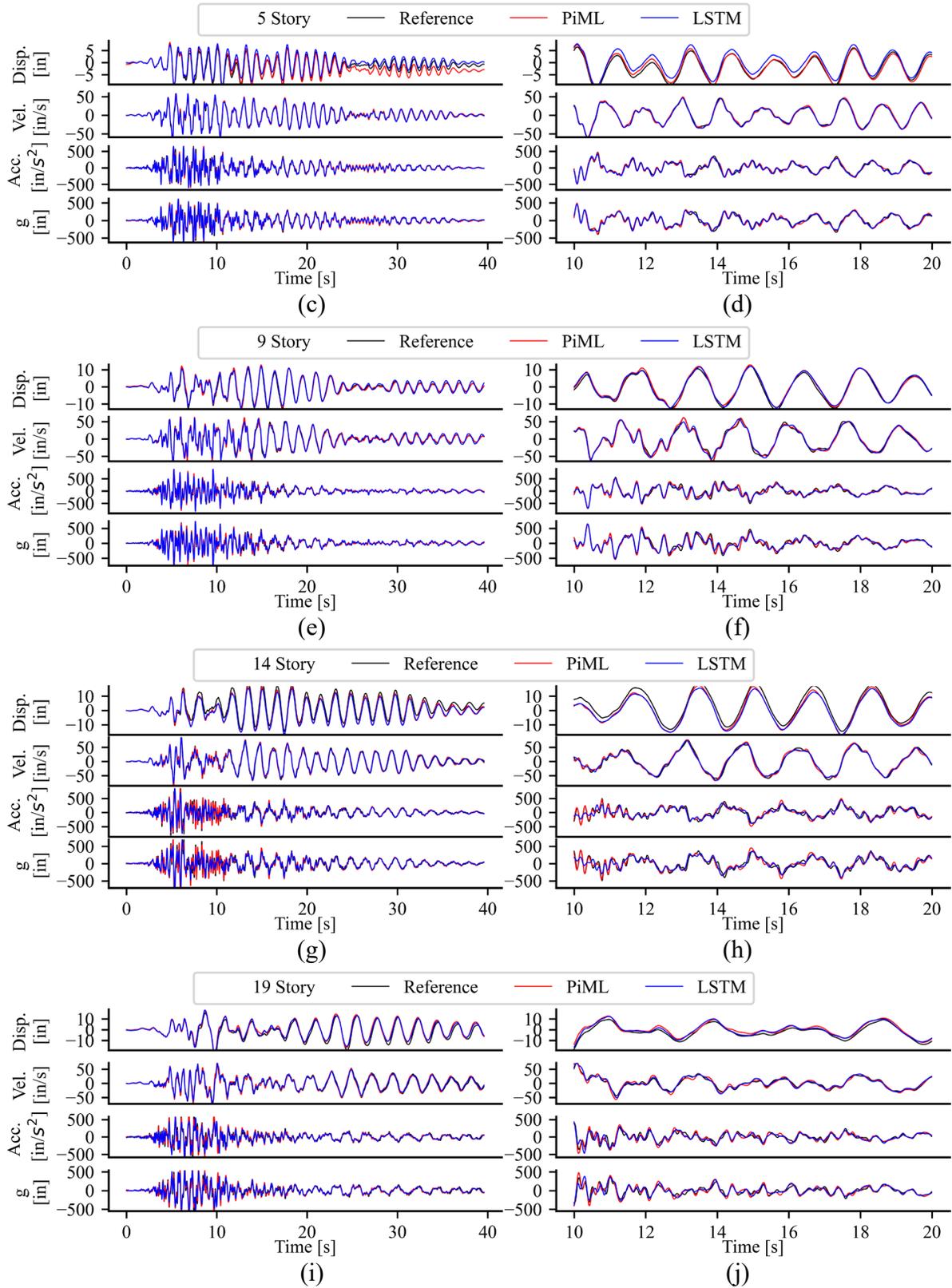

*Figure 7: Comparative outputs of response state variables from the PiML metamodel and LSTM network, showcasing: (a) a 1 story SMRF typical response; (b) zoomed-in signal of (a); (c) a 5 story SMRF's typical response; (d) an zoomed-in signal of (c);(e) a 9 story SMRF's typical response; (f) zoomed-in signal of (e); (g) a 14 story SMRF's typical response; (h) zoomed-in signal of (g); (i) a 19 story SMRF's typical response; and (j) zoomed-in signal of (i).*

*3.3 Evaluation and Validation*

To showcase the model's robustness, a validation approach was implemented where each model was trained on 10 different, randomly chosen subsets from the full set of data. This approach is similar to the widely recognized 10-fold cross-validation technique in ML, which mitigates risks of overfitting or underfitting that can arise from using a single test set [69]. In this approach, 100 data pairs are chosen from the full suite of 240 data pairs for training randomly 10 times. Each subset is utilized once for training, and the model's performance is evaluated on the remaining data, allowing for error calculation across various configurations. This strategy not only decreases the variance in performance assessment compared to a single hold-out test set but also provides a more comprehensive evaluation over diverse data subsets.

The performance of the models is evaluated using various metrics, designed to assess different aspects of model accuracy and reliability. A widely used measure for comparing two time-series is the root mean square error normalized by the range of the signal ($RMSE^{range}$). Normalizing by range is beneficial for comparing data with varying magnitudes. This is a common metric that can provide a single evaluation value for an entire suite of test and predicted time-series (Eqn. (25)). Additionally, the Pearson's correlation coefficient (CC) is selected to offer a single evaluation figure, providing insight into the correlation between the predicted and reference response time-series (Eqn. (26)).

$$RMSE^{range} = \frac{1}{l}\sum_{i}^{l} \sqrt{\frac{\left(y_l^{true} - y_l^{pred}\right)^2}{\max(y_l^{true}) - \min(y_l^{truth})}} \tag{25}$$

$$CC = \frac{1}{l}\sum_{i}^{l} \frac{\sum_{i}^{\tau}(y_{l,i}^{true} - \bar{y}_l^{true})(y_{l,i}^{pred} - \bar{y}_l^{pred})}{\sqrt{\sum_{i}^{\tau}\left(y_{l,i}^{true} - \bar{y}_l^{true}\right)^2}\sqrt{\sum_{i}^{\tau}\left(y_{l,i}^{pred} - \bar{y}_l^{pred}\right)^2}} \tag{26}$$

In the above equations, $y^{true}$ represents a specific state variable (e.g., displacement, velocity, acceleration, or restoring force), reference response, $y^{pred}$ is either the state variable response of the PiML or the LSTM network, $l$ is the number of ground motions in the test set (e.g., 140), $\tau$ is the sequence length of the signal, and a bar represents the mean of the signal. A box and whisker plot representation of the $RMSE^{range}$ and the CC of the testing data set for the 10 trials are shown in Figure 8 and Figure 9 respectively. For each metric and each structure, the PiML metamodel has lower mean errors with fewer outliers overall compared to the data-driven LSTM network.

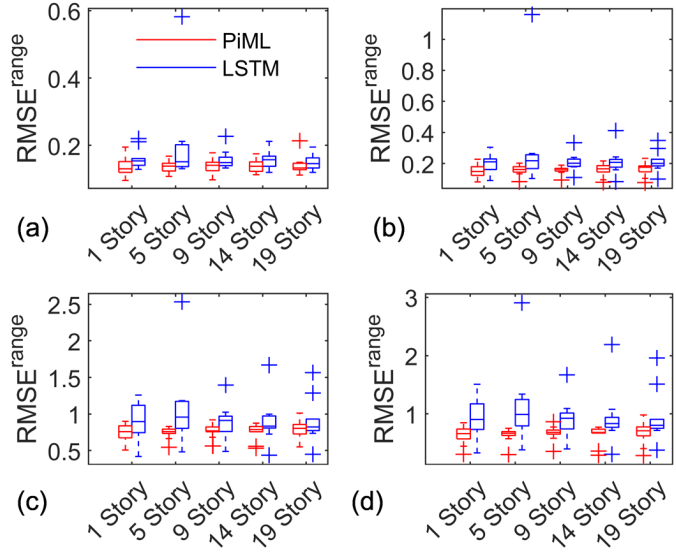

*Figure 8: A box-and-whisker plot for the RMSE normalized by range for 10 independent trials of the PiML vs. LSTM for (a) displacement error; (b) velocity error; (c) acceleration error; and (d) the restoring force error.*

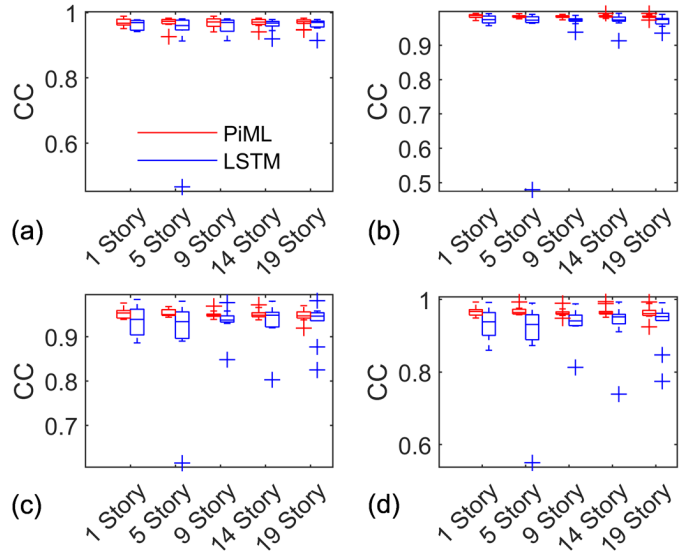

*Figure 9: A box-and-whisker plot for the correlation coefficient (CC) for 10 independent trials of the PiML vs. LSTM for (a) displacement; (b) velocity; (c) acceleration; and (d) the restoring force*

The peak error ($L_{peak}$) measures the difference between the peak values in the predicted and reference data, displaying the model's ability to accurately replicate the extreme values in the response series (e.g., peak story drift, etc.) (Eqn. (27)). The amplitude loss calculates the average of the amplitude differences over '$\tau$' time steps evaluating the model's precision in replicating the magnitude of fluctuations in the time-series across the duration of the signal (Eqn. (28)). The energy loss quantifies the relative difference in the integrated response over time between the predicted and original data (Eqn. (29)).

$$L_{peak} = \frac{max|y^{true}| - max|y^{pred}|}{max|y^{true}|} \quad (27)$$

$$A = \frac{1}{\tau} \sum_i^\tau \frac{|y_i^{true} - y_i^{pred}|}{|y_i^{true}|} \quad (28)$$

$$\mathcal{E} = \frac{\sum_i^\tau |y_i^{true}| - \sum_i^\tau |y_i^{pred}|}{\sum_i^\tau |y_i^{true}|} \quad (29)$$

One value is taken for each of these three-error metrics (e.g., $L_{peak}$, $A$, and $\mathcal{E}$) for all examples in the test set (e.g., 140 data pairs). These 140 error metric values are then converted into a cumulative distribution function (CDF) to show the probability that the metric is above or below a certain value. Individual CDFs for the 10 trials are show as thin lines for the PiML (red) and LSTM (blue) metamodels in Figure 10, Figure 11, and Figure 12. A thick line represents the mean of all CDFs for the PiML an LSTM metamodels. It is shown, for the majority of the structures and state variables, that the PiML outperforms the LSTM network in all error metrics.

Importantly, metrics such as the $RMSE^{range}$, peak loss, and amplitude loss focus on localized prediction quality, scrutinizing the model's accuracy at individual time steps. In contrast, the correlation coefficient and energy loss compare the global quality of the predictions, considering the entire response series.

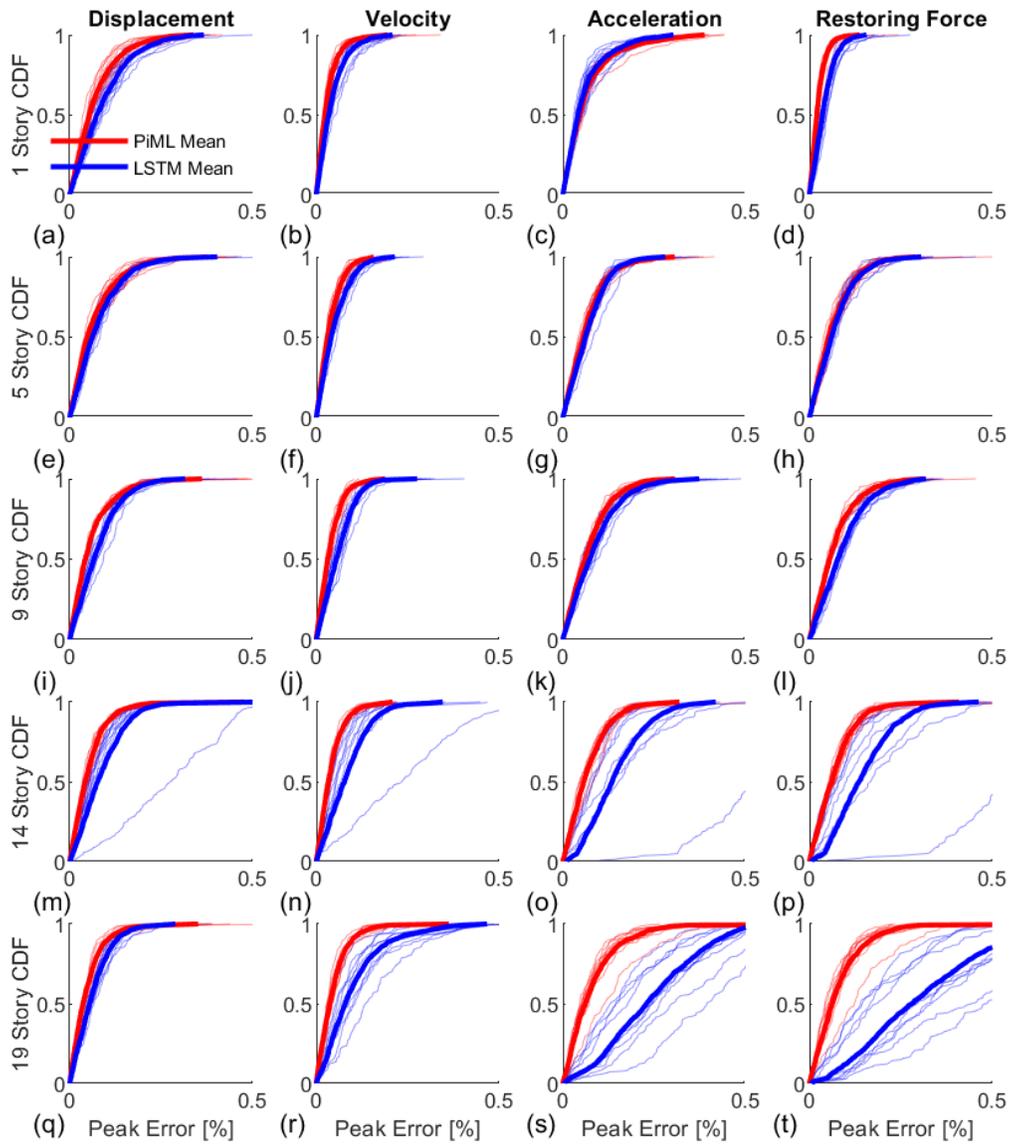

*Figure 10: Peak error for all frames and all state variables*

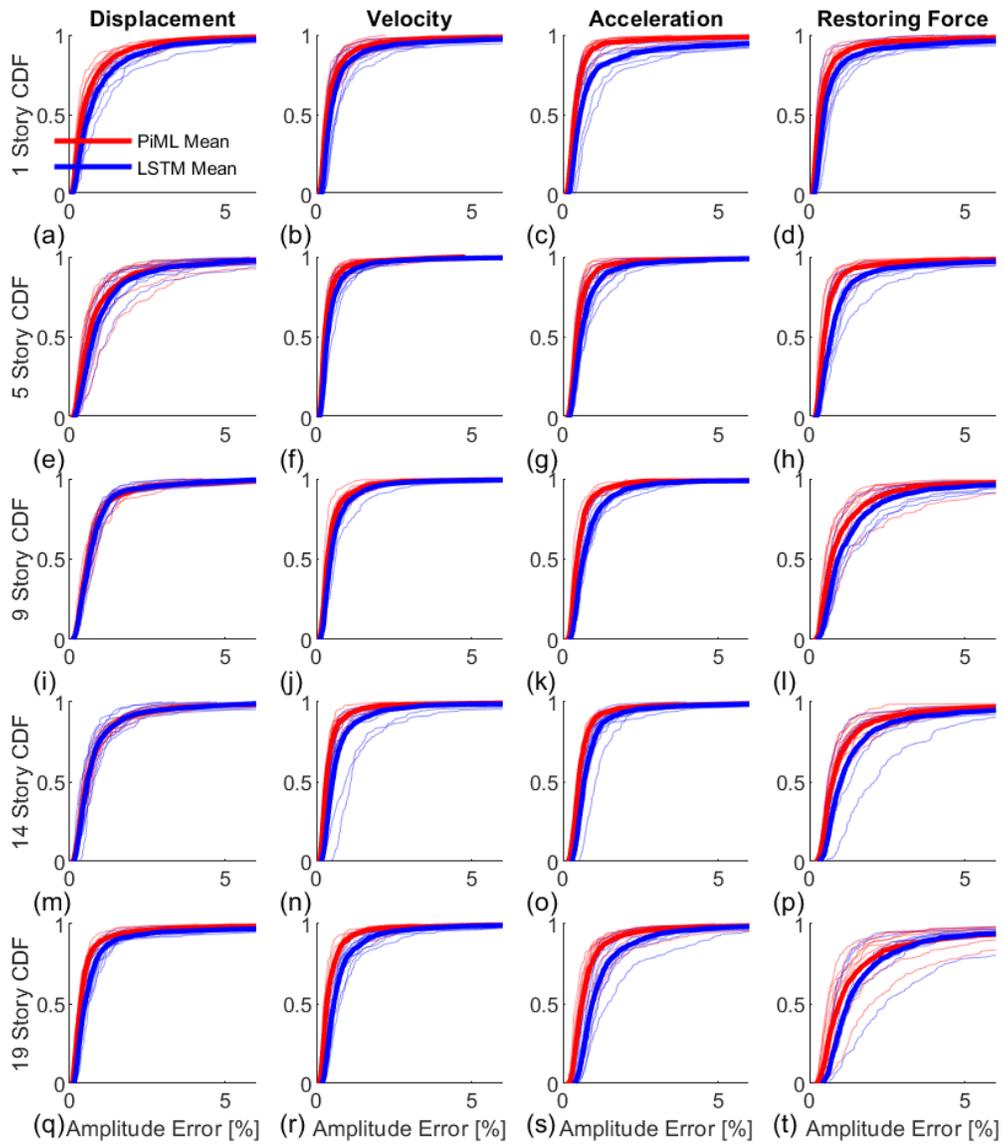

Figure: Amplitude error for all frames and all state variables

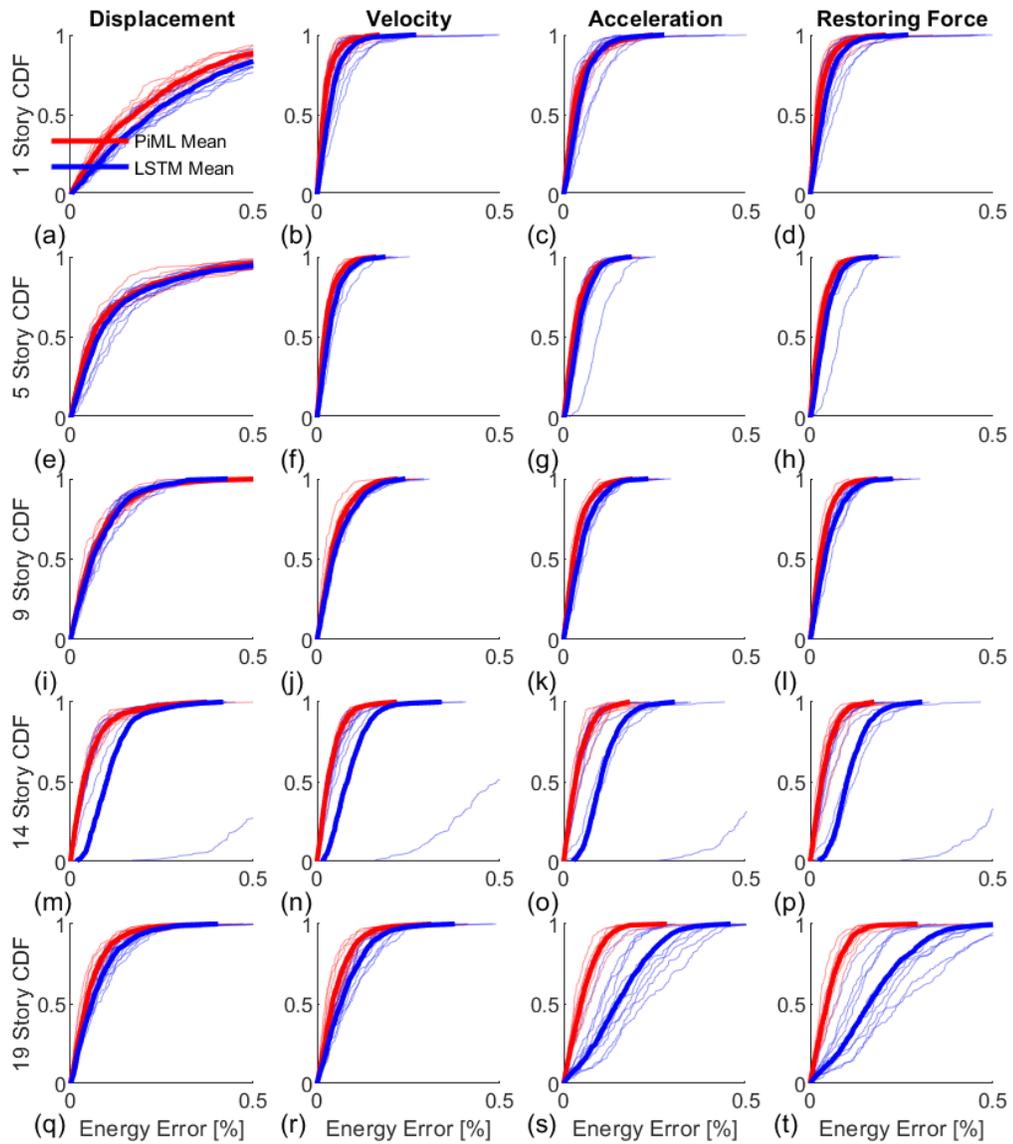

*Figure 11: Energy error for all frames and all state variables*

## 3.4 Future Work

The dataset of 621 SMRF was chosen to expand on the scope of the PiML metamodels. Transfer learning is currently being explored to transfer model parameters for one structure to another to save training effort and maintain accuracy for other structural models in the database. Additionally, various use case applications of this model, like incremental dynamic analysis, fragility analysis, and development of seismic performance factors, are being considered in our future work.

## 4. CONCLUSION

This research incorporates physical constraints of traditional earthquake engineering and structural dynamics with innovative elements of ML, creating a novel physics-informed deep learning approach for predicting nonlinear structural seismic responses. Particularly, the PiML metamodel is capable of mapping seismic input to detailed global response time-histories of response state variables, (e.g., $\mathbf{a_g} \xrightarrow{\text{metamodeling}} \mathbf{u_{tt}, u_t, u, g}$). The EOM and a differentiation filter are taken as constraints and incorporated in the ML network architecture to guide the model training to a physics-based solution space. Another distinguishing feature of the proposed framework is the application of the wavelet transformation, a method known to capture localized features and variation, capable of providing a flexible time-frequency deconstruction and reconstruction of the time-series input/output pairs. In this case, wavelets were particularly found to effectively handle the non-stationarity and irregularity in the residual drift of the displacement data, thereby enhancing the accuracy and nuance of PiML metamodels. In this way, the trained PiML metamodel can accurately capture structural dynamics even with limited training/validation data.

For demonstration purposes, the PiML metamodeling architecture is tested on steel moment resisting frame models downloaded from the Design-Safe CI data depot [56,57] and compared with a non-physics guided, data-driven LSTM network. For all five frame structures, the PiML outperforms the data-driven network in reconstruction in the state variable responses.

The PiML metamodel introduced in this work provides three key benefits. Firstly, it ensures clear interpretability by integrating physical principles, which not only lends meaning to the model's outputs but also enhances the understanding of its operations. Secondly, the model demonstrates superior generalizability and robust inference capabilities, attributed to the incorporation of embedded physics. This approach effectively constrains network outputs, mitigates overfitting, and diminishes the reliance on extensive training datasets, thereby bolstering model reliability for predictions. Lastly, the model boasts exceptional efficiency and accuracy in its performance, making it adept at handling datasets of limited richness without compromising on speed or precision.

## 5. ACKNOWLEDGMENTS

This material is based upon work supported by the National Science Foundation under Grant No. CMMI-2013067 and Northeastern University. The authors would like to thank James Fong for his assistance with this research.

## 6. DATA AVAILABILITY STATEMENT

All the source codes to reproduce the results in this study are available on GitHub at https://github.com (detailed URL will be provided after official publication of this paper).